\documentclass[prd,aps,preprintnumbers, showpacs, nofootinbib,superscriptaddress,
notitlepage
]{revtex4-1}
  \usepackage{latexsym,bm,amsmath,amssymb,amsfonts,color}
\newcommand{\beq}{\begin{eqnarray}}
\newcommand{\eeq}{\end{eqnarray}}

\newcommand{\nn}{\nonumber \\}

\newcommand{\Slash}[1]{{\ooalign{\hfil/\hfil\crcr$#1$}}}

\usepackage{graphicx}
\usepackage{amsmath,amsthm,amssymb}
\usepackage{color}

\begin{document}
\preprint{YITP--18--112}
\preprint{J-PARC TH--0136}

\title{Quark and gluon contributions to the QCD trace anomaly }

\author{Yoshitaka Hatta}

\affiliation{Physics Department, Brookhaven National Laboratory, Upton, New York 11973, USA}
\affiliation{Yukawa Institute for Theoretical Physics, Kyoto University, Kyoto 606-8502, Japan}

\author{Abha Rajan}
\affiliation{Physics Department, Brookhaven National Laboratory, Upton, New York 11973, USA}

\author{Kazuhiro Tanaka}
\affiliation{Department of Physics, Juntendo University, Inzai, Chiba 270-1695, Japan }
\affiliation{J-PARC Branch, KEK Theory Center, Institute of Particle and Nuclear Studies, KEK, 203-1, Shirakata, Tokai, Ibaraki, 31901106, Japan}


\begin{abstract}
We show that, in dimensional reguralization in the minimal subtraction scheme, the QCD trace anomaly can be unambiguously decomposed into two parts coming from the renormalized quark and gluon energy momentum tensors. We carry out this decomposition at the two-loop level.  
 The result can be used to constrain the renormalization group properties of the nucleon's twist-four gravitational form factor $\bar{C}_{q,g}$.  
\end{abstract}
\maketitle

\section{Introduction}
It is well known that the QCD Lagrangian is classically scale invariant if the small quark masses are neglected, but the invariance is broken at the quantum level. A common way to  express this phenomenon is to compute the trace of the QCD energy momentum tensor $T^{\mu\nu}$ 
\beq
T^\alpha_\alpha=m (1+\gamma_m)\bar{\psi}\psi+ \frac{\beta(g)}{2g}F^{\mu\nu}F_{\mu\nu}.  \label{111}
\eeq
In addition to the expected quark mass contribution, terms proportional to 
the beta-function $\beta$ and the mass anomalous dimension $\gamma_m$ appear. This is called the trace anomaly and is fundamentally important in QCD as it signals the generation of a nonperturbative mass scale.  In particular, its  nucleon matrix element  $\langle P|T^\alpha_\alpha|P\rangle$ (relative to the vacuum expectation value) is proportional to the nucleon mass squared (see (\ref{mass}) below).

In this paper, we address the following question. The energy momentum tensor consists of the quark part and the gluon part $T^{\mu\nu}=T_q^{\mu\nu} +T_g^{\mu\nu}$. Which part of the anomaly (\ref{111}) comes from $T^{\mu\nu}_q$ and the rest from $T^{\mu\nu}_g$? Until recently, there has not been enough motivation to ask this question. First of all, it is not clear {\it a priori} whether such a decomposition is well-defined,  and even if the answer is yes, it appears to be a purely conceptual problem without any phenomenological implications. More technically, while the total anomaly is renormalization group (RG) invariant, the individual terms  $T^{\alpha}_{q \alpha}$ and $T^{\alpha}_{g \alpha}$ are not. Furthermore, the decomposition may depend on the regularization scheme one is using to handle the UV divergences. 

However, in recent years the necessity to understand the QCD energy momentum tensor has intensified significantly. It has become a common practice to parametrize the `gravitational form factor' of hadrons separately for quarks and gluons $\langle P'|T^{\mu\nu}_{q,g}|P\rangle$ \cite{Ji:1996ek,Kumano:2017lhr,Polyakov:2018zvc,Tanaka:2018wea}.
Very importantly,  understanding the origin of the nucleon mass has emerged as one of the main objectives of  the future Electron-Ion Collider \cite{nas}, and  
 an independent experiment dedicated to this issue has been proposed at the Jefferson Laboratory \cite{Joosten:2018gyo}. Specifically, JLab proposes to measure the near-threshold photoproduction of $J/\psi$ in $ep$ scattering. It has been shown in \cite{Kharzeev:1998bz,Hatta:2018ina}   that the cross section of this process is sensitive to the $F^2$ part of the trace anomaly (\ref{111}). However, the extraction of the forward matrix element $\langle P|F^2|P\rangle$ from the experimental data is complicated by the fact that near the threshold, the momentum transfer $\Delta=P'-P$ is not negligible. In order to facilitate the extrapolation to $\Delta\to 0$, it is convenient to express $\langle P'|F^2|P\rangle$ in terms of the gravitational form factors $\langle P'|(T_{q,g})^\alpha_\alpha|P\rangle$.  In Ref.~\cite{Hatta:2018ina}, this relation has been worked out at the level of the bare operators. In this paper, we show that the relation gets modified once one considers the renormalized operators, and compute the correction to two loops in dimensional regularization in the minimal subtraction scheme.  This clarifies the relation between the bare and  renormalized operators $(T_{q,g})^\alpha_\alpha$. As an important application of our result, we  elucidate  the renormalization group property of the twist-four gravitational form factor $\bar{C}_{q,g}$.

\section{The trace anomaly}

Our starting point is the gauge-invariant, symmetric QCD energy momentum tensor which is given by
\beq
T^{\mu\nu}=-F^{\mu\lambda}F^\nu_{\ \lambda} + \frac{\eta^{\mu\nu}}{4}F^2 + i\bar{\psi}\gamma^{(\mu}\overleftrightarrow{D}^{\nu)}\psi  = T^{\mu\nu}_g + T^{\mu\nu}_q,
\eeq
 where $D^\mu=\partial^\mu+ig A^\mu$, $A^{(\mu}B^{\nu)}\equiv \frac{A^\mu B^\nu+A^\nu B^\mu}{2}$ and $\overleftrightarrow{D}^\mu \equiv \frac{D^\mu -\overleftarrow{D}^\mu}{2}$.  We have neglected the ghost and gauge fixing terms as they do not affect our final results. 
 In the last equality, we decomposed the total energy momentum tensor into the gluon and quark parts as 
$T_g^{\mu\nu}= -F^{\mu\lambda}F^\nu_{\ \lambda} + \frac{\eta^{\mu\nu}}{4}F^2$ and $T^{\mu\nu}_q= i\bar{\psi}\gamma^{(\mu}\overleftrightarrow{D}^{\nu)}\psi$. $T^{\mu\nu}$ is conserved and therefore it is a finite, scale-independent operator. However, $T_g^{\mu\nu}$ and $T_q^{\mu\nu}$ are not conserved separately and are subject to regularization and renormalization. 
 
 We work in  $d=4-2\epsilon$ spacetime dimensions. Let us decompose $T^{\mu\nu}$ into the traceless $\bar{T}^{\mu\nu}$ and trace $\hat{T}^{\mu\nu}$ parts. 
\beq
T^{\mu\nu}=\left( T^{\mu\nu}-\frac{\eta^{\mu\nu}}{d} T^{\alpha}_\alpha \right)+ \frac{\eta^{\mu\nu}}{d}T^\alpha_\alpha \equiv \bar{T}^{\mu\nu} + \hat{T}^{\mu\nu},
\label{trace}
\eeq
where 
\beq
T^\alpha_\alpha = -2\epsilon \frac{F^2}{4} + \bar{\psi} i\overleftrightarrow{\Slash D}\psi = -2\epsilon \frac{F^2}{4} + m\bar{\psi} \psi. \label{an}
\eeq
The second equality follows from the equation of motion. ($m$ is the quark mass.) 
The corresponding decomposition for the {\it bare} operators $T^{\mu\nu}_{q,g}$ is 
\beq
T^{\mu\nu}_g=\left(-F^{\mu\lambda}F^\nu_{\ \lambda} +\frac{\eta^{\mu\nu}}{d}F^2 \right)+ \frac{\eta^{\mu\nu}}{d} \frac{-2\epsilon}{4}F^2,
\eeq
\beq
T_q^{\mu\nu}= \left(i\bar{\psi}\gamma^{(\mu}\overleftrightarrow{D}^{\nu)}\psi -\frac{\eta^{\mu\nu}}{d} i\bar{\psi} \overleftrightarrow{\Slash D} \psi\right) +\frac{\eta^{\mu\nu}}{d}i\bar{\psi}\overleftrightarrow{\Slash D}\psi. 
\eeq
The operator $F^2$ is divergent and has to be regularized. We shall use the (modified) minimal subtraction scheme, and in this scheme the renormalization of $F^2$ has been well understood in the literature \cite{Nielsen:1977sy,Tarrach:1981bi}.  Denoting renormalized operators with a sub- or super-script $R$,  one finds $m\bar{\psi}\psi = (m\bar{\psi} \psi)_R$ and
\beq
-2\epsilon  \frac{F^2}{4}  = \frac{\beta(g_R)}{2g_R}(F^2)_R +  \gamma_m (m \bar{\psi}\psi)_R
\eeq
 where $\beta$ is the QCD beta-function $\beta(g_R) = \frac{\partial g_R(\mu)}{\partial \ln \mu}$ and $\gamma_m(g_R)= -\frac{1}{m_R} \frac{\partial m_R(\mu)}{\partial \ln \mu}$ is the mass anomalous dimension. 
We thus arrive at the standard result
\beq
T^{\alpha}_\alpha = \frac{\beta_R}{2g_R}(F^2)_R + (1+\gamma^R_m)(m\bar{\psi} \psi)_R.
\eeq
The above derivation makes it clear that, in dimensional regularization, the anomaly entirely comes from the bare gluon energy momentum tensor, while the bare quark energy momentum tensor only contributes to the mass term 
\beq
T^{\alpha}_{g\alpha} &=& \frac{\beta_R}{2g_R}(F^2)_R + \gamma^R_m (m\bar{\psi} \psi)_R , \label{1}\\
T_{q\alpha}^\alpha   &=& (m\bar{\psi} \psi)_R. \label{2}
\eeq
Noting that the right hand sides of (\ref{1}) and (\ref{2}) are both renormalization-group (RG) invariant, one can also write 
\beq
T^{\alpha}_{g\alpha} &=& \frac{\beta}{2g}F^2 + \gamma_m m\bar{\psi} \psi, \\
T_{q\alpha}^\alpha    &=& m\bar{\psi} \psi , \label{alpha}
\eeq
 where all the quantities and fields are bare. Note that such a clean separation of the trace anomaly into the quark and gluon parts may not be unambiguously done in other regularization schemes. For example, in the Pauli-Villars regularization, the anomaly comes from the energy-momentum tensor of the massive regulator field which is neither $T_q$ nor $T_g$.\footnote{We thank M. Polyakov for providing this argument. } 
 The goal of this paper is to derive the corresponding  formulas for the {\it renormalized}  operators $(T_{qR})^\alpha_{\alpha}$ and $(T_{gR})^\alpha_\alpha$.

  \section{Nucleon gravitational form factors}
  
 The scale-dependence of $T_{q,g}^R$ essentially determins the scale dependence of the so-called gravitational form factors.  The nonforward nucleon matrix element of $T^{\mu\nu}_{q,g}$ and $(T^{\mu\nu}_{q,g})_R$ can be parametrized as  
\beq
\langle P'|T^{\mu\nu}_{q,g}|P\rangle = \bar{u}(P')\Bigl[ A_{q,g}\gamma^{(\mu}\bar{P}^{\nu)} + B_{q,g}\frac{\bar{P}^{(\mu}i\sigma^{\nu)\alpha}\Delta_\alpha}{2M} + C_{q,g}\frac{\Delta^\mu\Delta^\nu-\eta^{\mu\nu}\Delta^2}{M} + \bar{C}_{q,g}M\eta^{\mu\nu} \Bigr] u(P),  \label{grav}
\eeq
\beq
\langle P'|(T^{\mu\nu}_{q,g})_R|P\rangle = \bar{u}(P')\Bigl[ A^R_{q,g}\gamma^{(\mu}\bar{P}^{\nu)} + B^R_{q,g}\frac{\bar{P}^{(\mu}i\sigma^{\nu)\alpha}\Delta_\alpha}{2M} + C^R_{q,g}\frac{\Delta^\mu\Delta^\nu-\eta^{\mu\nu}\Delta^2}{M} + \bar{C}^R_{q,g}M\eta^{\mu\nu} \Bigr] u(P) \nn \label{t}
\eeq
where $\Delta^\mu=P'^\mu-P^\mu$ is the momentum transfer and $\bar{P}^\mu\equiv \frac{P^\mu+P'^\mu}{2}$. $M$ is the nucleon mass. The conservation of the energy momentum tensor implies that $A^{(R)}_q+A^{(R)}_g=1$ at $\Delta=0$ and $\bar{C}^{(R)}_q=-\bar{C}^{(R)}_g$ for all values of $\Delta$. The renormalized operators are given by
\beq
T_{gR}^{\mu\nu} = -(F^{\mu\lambda}F^\nu_{\ \lambda })_R + \frac{\eta^{\mu\nu}}{4}(F^2)_R,  \qquad T_{qR}^{\mu\nu} =i (\bar{\psi}\gamma^{(\mu}\overleftrightarrow{D}^{\nu)}\psi)_R  ,
\eeq
and the form factors $A_R,B_R,C_R,\bar{C}_R$ are renormalized at scale $\mu$.  Naively, since $(F^2)_R$ is now a finite operator, one might think that $(T_{gR})^\alpha_\alpha=0$. However, this is not the case, because renormalization and the trace operation do not commute in dimensional regularization (see, e.g., \cite{Collins:1984xc,Suzuki:2013gza}).
Taking the trace as well as the forward limit, we find 
\beq
\langle P|(T_{g})^\alpha_\alpha|P\rangle&=& \langle P|\left( \frac{\beta}{2g}F^2 + \gamma_m m\bar{\psi} \psi \right)|P\rangle= 2M^2(A_g+d\bar{C}_g), \label{beta}\\
\langle P|(T_{q})^{\alpha}_{\alpha}|P\rangle &=&\langle P| m\bar{\psi}\psi|P\rangle= 2M^2(A_q+d\bar{C}_q), \label{we}
\eeq
and
\beq
\langle P|(T_{gR})^\alpha_\alpha(\mu)|P\rangle = 2M^2(A^R_g(\mu)+4\bar{C}^R_g(\mu)), \label{qan} \\
\langle P|(T_{qR})^{\alpha}_{\alpha}(\mu)|P\rangle = 2M^2(A^R_q(\mu)+4\bar{C}^R_q(\mu)).
\eeq
The mass of the nucleon is given by 
\beq
2M^2=\langle P| \left( \frac{\beta}{2g}F^2 + (1+\gamma_m) m\bar{\psi} \psi \right)|P\rangle=\langle P| \left( \frac{\beta_R}{2g_R}(F^2)_R + (1+\gamma^R_m) (m\bar{\psi} \psi)_R \right)|P\rangle.
\label{mass}
\eeq
Note that, in the chiral limit, $\bar{C}_q=-\frac{1}{4}A_q$ \cite{Ji:1997gm}. As suggested in \cite{Ji:1997gm}, this  relation does not hold for the renormalized quantities.

The $\mu$-dependence of $A_{q,g}^R(\mu)$ is well known. Since $A_{q,g}$ are the matrix elements of the twist-two, spin-2 quark and gluon operators, their evolution is closed under evolution. To one-loop order, one finds
\beq
  \frac{\partial}{\partial \ln \mu} \begin{pmatrix} A^R_q \\ A^R_g \end{pmatrix}
=  \frac{\alpha_s}{4\pi} \begin{pmatrix} -\frac{16}{3}C_F & \frac{4n_f}{3} \\ \frac{16}{3}C_F & -\frac{4n_f}{3} \end{pmatrix}   \begin{pmatrix} A^R_q \\ A^R_g\end{pmatrix},  \label{one}
\eeq
 where $C_F=\frac{N_c^2-1}{2N_c}=\frac{4}{3}$. [To simplify the notation, in the following we write $\alpha_s^R=\frac{g_R^2}{4\pi}\equiv \alpha_s$ for the renormalized coupling.]
On the other hand, the $\mu$-dependence of $\bar{C}_{q,g}$ can be obtained as follows. First one uses the identity \cite{Braun:2004vf,Tanaka:2018wea} 
\beq
\partial_\nu T^{\mu\nu}_q = \bar{\psi} gF^{\mu\nu}\gamma_\nu \psi,  
 \label{phys}
\eeq
where the terms which vanish due to the equation of motion have been neglected. Similarly,
\beq
\partial_\nu T^{\mu\nu}_g = F_{\nu }^{\ \mu} D_\alpha  F^{\alpha\nu}. \label{eq2}
\eeq
Note that (\ref{phys}) and (\ref{eq2}) are compatible with the condition $\partial_\nu (T^{\mu\nu}_q+T^{\mu\nu}_g)=0$ thanks to the equation of motion $D_\alpha F^{\alpha\nu}=g\bar{\psi}\gamma^\nu \psi$.  
Taking the matrix element of (\ref{phys}), one finds
\beq
\langle P'|g\bar{\psi}F^{\mu\nu}\gamma_\nu \psi |P\rangle =iM \Delta^\mu \bar{C}_q \bar{u}(P')u(P), \label{mat1} \\
\langle P'|  F_{\nu }^{\ \mu} D_\alpha F^{\alpha\nu}  |P\rangle =iM \Delta^\mu \bar{C}_g \bar{u}(P')u(P). \label{mat2}
\eeq
Therefore, the $\mu$-dependence of $\bar{C}_{q,g}$ is governed by the anomalous dimension of the twist-4 operators $g\bar{\psi}F^{\mu\nu}\gamma_\nu \psi$ and $F_{\nu }^{\ \mu} D_\alpha F^{\alpha\nu}$. The latter can be computed either directly, using the well-established techniques in the literature \cite{Shuryak:1981kj,Bukhvostov:1984as,Balitsky:1987bk,Kodaira:1996md,Kodaira:1997ig,Kodaira:1998jn,Nishikawa:2011qk}, or simply by noticing that it must coincide with the anomalous dimension of $T^{\mu\nu}_q$ by virtue of the identity (\ref{phys}). 
One finds, in the chiral limit $m=0$ \cite{Polyakov:2018exb}, 
\beq
\frac{\partial}{\partial \ln \mu} \bar{C}^R_q =- \frac{\alpha_s}{4\pi} \left(\frac{16}{3}C_F+\frac{4n_f}{3}\right)\bar{C}^R_q, \label{pol} 
\eeq
from which one would conclude that $\bar{C}^R_{q,g}(\mu)\to 0$ as $\mu\to \infty$ (in this chiral limit). 
We have computed the correction due to the quark mass with the result 
\beq
 \frac{\partial}{\partial \ln \mu} (g\bar{\psi}F^{\mu\nu}\gamma_\nu \psi)_R =  - \frac{\alpha_s}{4\pi} \left(\frac{16}{3}C_F+\frac{4n_f}{3}\right)  (g\bar{\psi}F^{\mu\nu}\gamma_\nu \psi)_R+
 \frac{4C_F}{3}\frac{ \alpha_s}{4\pi} \partial^\mu (m \bar{\psi}\psi)_R. \label{tot}
\eeq
 This implies that, in the forward limit $\Delta\to 0$, 
\beq
\frac{\partial}{\partial \ln \mu} \bar{C}^R_q =- \frac{\alpha_s}{4\pi} \left(\frac{16}{3}C_F+\frac{4n_f}{3}\right)\bar{C}^R_q + \frac{\alpha_s}{4\pi} \frac{4C_F}{3} \frac{1}{2M^2}\langle P |(m\bar{\psi}\psi)_R |P\rangle. \label{full}
\eeq

We note that although (\ref{mat1}) and (\ref{mat2}) make sense  only at nonzero momentum transfer $\Delta \neq 0$, after removing the common factor $\Delta^\mu$ the  limit $\Delta\to 0$ can be safely taken  to arrive at (\ref{full}). However,  Eq.~(\ref{full}) is actually  problematic. One would expect that the value of $\bar{C}_{q,g}$ should be at least partly constrained by the trace anomaly, but  (\ref{full}) appears to be insensitive to it. As we shall see in the next section, one has to include certain two-loop contributions  in order to obtain the correct asymptotic limit of $\bar{C}^R$.

 \section{One-loop renormalization of $T^R_{q,g}$}

Since the right hand sides of (\ref{beta}) and (\ref{we}) are both renormalization group invariant, one may naively think that $A_{q,g}+4\bar{C}_{q,g}$ is invariant under renormalization, i.e., $A_{q,g}+4\bar{C}_{q,g}=A_{q,g}^R(\mu)+4\bar{C}^R_{q,g}(\mu)$. However, this is not the case. In this section we show that this quantity receives a finite renormalization. 

For notational symplicity, let us write
\beq
&&O_1=-F^{\mu\lambda}F^{\nu}_{\ \lambda}, \\
&&O_2= \eta^{\mu\nu}F^2,\\
&&O_3=i\bar{\psi}\gamma^{(\mu}\overleftrightarrow{D}^{\nu)} \psi, \\
&&O_4=\eta^{\mu\nu}m\bar{\psi}\psi.
\eeq
Then the energy momentum tensor is
\beq
T^{\mu\nu}=O_1+\frac{O_2}{4}+O_3.
\eeq
We introduce the renormalization constants as\footnote{The most general formula includes the mixing with the equation-motion operators as well as the BRST-exact operators. However, they do not affect our final result because their matrix elements in a physical state vanish (see e.g., \cite{Collins:1984xc,Kodaira:1997ig,Kodaira:1998jn}).}
\beq
O_1^R&=&Z_T O_1 + Z_M O_2 + Z_L O_3 + Z_S O_4, \\
O_2^R&=&Z_F O_2+Z_C O_4, \label{o2ren}\\
O_3^R&=&Z_\psi O_3 +Z_KO_4+ Z_Q O_1 +Z_B O_2,  \label{pre} \\
O_4^R&=& O_4. \label{o4ren}
\eeq
To one-loop order \cite{Tarrach:1981bi}
\beq
Z_F&=&1 - \frac{\alpha_s}{2\pi }\beta_0\frac{1}{2\epsilon} \\
Z_C&=&  4\gamma_m \frac{1}{2\epsilon}
\eeq
 where $\beta_0=\frac{11}{3}C_A-\frac{2n_f}{3}$ with $C_A=N_c=3$ is the first coefficient of the beta-function $\beta(g)=-\beta_0 \frac{g^3}{16\pi^2}+\cdots$. 
 The mass anomalous dimension is given by $\gamma_m=\frac{3C_F\alpha_s}{2\pi}$ to this order.   
From the condition $T^{\mu\nu}=T^{\mu\nu}_R$, we get the following relations
\beq
&&Z_T+Z_Q=1,  \label{11} \\
&& Z_M+\frac{Z_F}{4}+Z_B=\frac{1}{4},\\
&& Z_L +Z_\psi= 1, \\
&& Z_S+\frac{Z_C}{4}+Z_K=0. \label{44}
\eeq
Moreover, the twist-two operators $O_1-({\rm trace})$ and $O_3-({\rm trace})$ are mixed among themselves under renormalization. The traceless gluonic twist-two operator is 
\beq
\widetilde{O}_1^{R}= O_1^{R}+\frac{\eta^{\mu\nu}}{d} \eta_{\alpha\beta}[F^{\alpha\lambda}F^\beta_{\ \lambda}]^R.
\eeq
Let us write 
\beq
 \eta_{\alpha\beta}[F^{\alpha\lambda}F^\beta_\lambda]^R =  \left(1-\frac{\beta}{2g}+x\right) (F^2)_R + (-\gamma_m+y)(m\bar{\psi}\psi)_R, \label{ei}
 \eeq
 where we have taken into account the fact that the trace operation and renormalization do not commute and parameterized the possible anomalous terms by the unknown constants  $x,y={\cal O}(\alpha_s)$. Note that $\frac{\beta}{2g}=-\frac{\alpha_s}{8\pi}(\frac{11}{3}C_A-\frac{2n_f}{3})$ to this order.  
Then the twist-two quark operator becomes\footnote{We have parametrized (\ref{ei}) and (\ref{ee}) such that their trace parts reproduce the total anomaly  (\ref{111}).   However, this is actually not necessary. The constraints (\ref{co1}) and (\ref{co2}) are strong enough that they completely determine the anomaly term. That is, we may introduce two new unknown parameters  for the coefficient of $O_{2,4}^R$ in (\ref{ee}) and still obtain the same result.  This is true also at two-loop to be discussed in the next section.}       
 \beq
 \widetilde{O}_3^{R}=O_3^{R}-\frac{x}{d}O_2^{R}-\frac{1+y}{d}O_4^R.  \label{ee}
 \eeq
The renormalization relation is (cf., (\ref{one})),  
\beq
&& O_1^R +\left(1-\frac{\beta}{2g}+x\right)\frac{O_2^R}{d} + (-\gamma_m+y)\frac{O_4^R}{d} \nn 
&& \qquad \qquad = \left(1+\frac{\alpha_s}{4\pi\epsilon}\frac{2n_f}{3}\right)\left(O_1+\frac{O_2}{d} \right) -\frac{\alpha_s}{4\pi\epsilon}\frac{8C_F}{3} \left( O_3 -\frac{O_4}{d}\right), 
\label{co1} \\
&&O_3^R -x \frac{O_2^R}{d}  -\frac{1+y}{d}O_4^R= \left(1+\frac{\alpha_s}{4\pi\epsilon} \frac{8C_F}{3}\right) \left(O_3-\frac{O_4}{d}\right) -\frac{\alpha_s}{4\pi\epsilon}\frac{2n_f}{3}  \left(O_1+\frac{O_2}{d} \right). \label{co2}
\eeq
From these two equations, we find 
\beq
&&Z_\psi = 1+\frac{\alpha_s}{4\pi} \frac{8C_F}{3\epsilon},  \\
&&Z_Q = -\frac{\alpha_s}{4\pi} \frac{2n_f}{3\epsilon}, \\
&& Z_B-\frac{x}{d}Z_F=-\frac{\alpha_s}{4\pi} \frac{2n_f}{3d \epsilon }, \\
&&Z_T = 1+\frac{\alpha_s}{4\pi}\frac{2n_f}{3\epsilon},\\
&& Z_L= -\frac{\alpha_s}{4\pi} \frac{8C_F}{3\epsilon }, \\
&& Z_M +\frac{1}{d}\left(1+x-\frac{\beta}{2g}\right) Z_F=\frac{1}{d}\left(1+\frac{\alpha_s}{4\pi}\frac{2n_f}{3\epsilon}\right), \\
&& dZ_K=xZ_C + 1+y-\left(1+\frac{\alpha_s}{4\pi} \frac{8C_F}{3\epsilon} \right), \\
&& Z_S + \left(1-\frac{\beta}{2g}+x\right)\frac{Z_C}{d}+ \frac{-\gamma_m+y}{d}=\frac{\alpha_s}{4\pi}\frac{8C_F}{3d\epsilon }.
\eeq
Combining these relations with (\ref{11})-(\ref{44}), we obtain the unique solution to this set of equations
\beq
Z_B&=& \frac{\alpha_s}{4\pi}\left(-\frac{n_f}{6\epsilon}\right) ,\\
Z_M &=& \frac{\alpha_s}{4\pi} \frac{11C_F}{12\epsilon}, \\
Z_K&=& \frac{\alpha_s}{4\pi} \left(-\frac{2C_F}{3\epsilon}\right),\\
Z_S&=& \frac{\alpha_s}{4\pi}\left(-\frac{7C_F}{3\epsilon}\right),
\eeq
and 
\beq
x=\frac{\alpha_s}{4\pi}\frac{n_f}{3}, \qquad y=\frac{\alpha_s}{4\pi}\frac{4C_F}{3}.
\eeq
We thus arrive at
\beq
\eta_{\mu\nu}T^{\mu\nu}_{gR} &=&
 \frac{\alpha_s}{4\pi}\left(-\frac{11C_A}{6} (F^2)_R + \frac{14C_F}{3}(m\bar{\psi}\psi)_R\right), \label{tgr}\\
\eta_{\mu\nu}T^{\mu\nu}_{qR}&=&
 (m\bar{\psi}\psi)_R+ \frac{\alpha_s}{4\pi} \left(\frac{n_f}{3}(F^2)_R + \frac{4C_F}{3}(m\bar{\psi}\psi)_R\right).
\eeq
In terms of the matrix element,
\beq
A_g^R(\mu)+4\bar{C}_g^R(\mu)&=&  \frac{1}{2M^2} \langle P|  \frac{\alpha_s}{4\pi}\left(-\frac{11C_A}{6} (F^2)_R + \frac{14C_F}{3}(m\bar{\psi}\psi)_R\right)   |P\rangle, \label{con}
\\
A_q^R(\mu)+4\bar{C}_q^R(\mu)&=&  \frac{1}{2M^2} \langle P| \left[ (m\bar{\psi}\psi)_R+ \frac{\alpha_s}{4\pi} \left(\frac{n_f}{3}(F^2)_R + \frac{4C_F}{3}(m\bar{\psi}\psi)_R\right)\right]  |P\rangle.
\label{cons}
\eeq
Taking the $\partial_\mu$-derivative of (\ref{pre}), we find 
\beq
(\bar{\psi}gF^{\mu\nu}\gamma_\nu \psi)_R = (Z_\psi-Z_Q)
\bar{\psi}gF^{\mu\nu}\gamma_\nu \psi
+ Z_K\partial^\mu (m\bar{\psi}\psi)
+\left(Z_B-\frac{Z_Q}{4}\right)\partial^\mu F^2
.
\label{rge3b}
\eeq
Noting that $Z_B-\frac{Z_Q}{4}=0$ to one-loop, we see that  the relation (\ref{tot}) is reproduced. 
On the other hand,  (\ref{pre}) can be written as  
\beq
T_{qR}^{\mu\nu} = T^{\mu\nu}_q + \frac{\alpha_s}{2\pi}\frac{1}{2\epsilon} \frac{8C_F}{3}\left(T^{\mu\nu}_q-\frac{\eta^{\mu\nu}}{4}m\bar{\psi}\psi\right) -\frac{\alpha_s}{2\pi} \frac{2n_f}{3}\frac{1}{2\epsilon} T^{\mu\nu}_g.
\eeq
Taking the trace and the forward matrix element, we get
\beq
A_q^R(\mu)+4\bar{C}_q^R(\mu) = A_q+d\bar{C}_q+\frac{1}{2M^2}\langle P| \frac{\alpha_s}{4\pi} \left(\frac{4C_F}{3} m\bar{\psi}\psi + \frac{n_f}{3}F^2\right)|P\rangle 
  . \label{50}
\eeq
In the last term we may replace $F^2\to (F^2)_R$ since the difference is ${\cal O}(\alpha_s^2)$. Then (\ref{50}) becomes  consistent with (\ref{cons}) after taking into account (\ref{we}).

Eq.~(\ref{50}) shows that $A_q^R+4\bar{C}_q^R$ is RG-invariant to ${\cal O}(\alpha_s)$, but it gets a finite renormalization with respect to the bare quantities.\footnote{Using (\ref{fir}), one may write 
\beq
A_q+d\bar{C}_q=A_q+4\bar{C}_q-\frac{\alpha_s}{4\pi} \left(\frac{4C_F}{3}A_q^R -\frac{n_f}{3}A_g^R\right). 
\eeq
But the combination $A_q+d\bar{C}_q$ is more useful as it is directly related to the mass term as in (\ref{we}). 
}
  From the RG equation $\frac{\partial}{\partial \ln \mu} (A_{q,g}+d\bar{C}_{q,g})=0$, we can deduce that
\beq
\frac{\partial}{\partial \ln \mu} \bar{C}^R_q = \frac{\alpha_s}{4\pi} \left(\frac{4C_F}{3} A^R_q - \frac{n_f}{3} A^R_g\right), \label{fir} \\
\frac{\partial}{\partial \ln \mu} \bar{C}^R_g = -\frac{\alpha_s}{4\pi}  \left(\frac{4C_F}{3} A^R_q - \frac{n_f}{3} A^R_g\right).
 \label{ano}
\eeq
This can be rewritten as 
\beq
\frac{\partial \bar{C}^R_q}{\partial \ln \mu}  &=& \frac{\alpha_s}{4\pi} \left(\frac{4C_F}{3}+\frac{n_f}{3}\right)(A_q^R -A_q^R(\infty)) \label{ban} \\
&=&- \frac{\alpha_s}{4\pi} \left(\frac{16C_F}{3}+\frac{4n_f}{3}\right)(\bar{C}_q^R -\bar{C}_q^R(\infty)) \nn 
&=& - \frac{\alpha_s}{4\pi} \left(\frac{16C_F}{3}+\frac{4n_f}{3}\right)\bar{C}_q^R + \frac{\alpha_s}{4\pi} \left[\frac{4C_F}{3} \frac{\langle P|(m\bar{\psi}\psi)_R|P\rangle}{2M^2} + \frac{n_f}{3}\left(\frac{\langle P|(m\bar{\psi}\psi)_R|P\rangle}{2M^2} -1\right) \right] + {\cal O}(\alpha_s^2),
\nonumber
\eeq
where  $A_q^R(\infty)=\frac{n_f}{4C_F+n_f}$ and in the second line we used $A_q^R(\mu)+4\bar{C}_q^R(\mu)=A_q^R(\infty)+4\bar{C}^R_q(\infty)$ to this order. In the third line, we used (\ref{cons}). 
Formally, (\ref{ban}) is consistent with (\ref{full}) to ${\cal O}(\alpha_s)$ accuracy because $\frac{\langle P|m\bar{\psi}\psi|P\rangle}{2M^2} -1 ={\cal O}(\alpha_s)$ due to the relation   
(\ref{mass}). However, the perturbative result (\ref{full})  misses the fact that the trace anomaly converts naively ${\cal O}(\alpha_s)$ terms $\alpha_s F^2$,  $\alpha_s m \bar{\psi}\psi$ into ${\cal O}(1)$ quantities.  The asymptotic limit of $\bar{C}^R_q(\mu)$ in the chiral limit in this one-loop approximation can be directly read off from (\ref{cons})  
\beq
\bar{C}^R_q(\infty) &=& \frac{1}{4} \left( -\frac{n_f}{4C_F+n_f} + \frac{1}{2M^2}\langle P| \frac{\alpha_s}{4\pi} \frac{n_f}{3} (F^2)_R |P\rangle \right) \nn
&=& - \frac{1}{4} \left( \frac{n_f}{4C_F+n_f} + \frac{2n_f}{3\beta_0}\right).
\label{asymptc}
\eeq
Numerically, $\bar{C}^R_q(\infty)\approx -0.146$ ($n_f=3$) and $\bar{C}_q^R(\infty)\approx -0.103$ ($n_f=2$). This is an order of magnitude larger than and has an opposite sign from the result of \cite{Polyakov:2018exb}.  While the two results are not necessarily inconsisntent, as they are obtained at different scales, a more detailed study is needed to clarify this issue (see (\ref{nu1}) and (\ref{nu2}) below).

\section{Renormalization at two-loop}

It is straightforward to generalize the result of the previous section to two-loop. 
The beta-function and the mass anomalous dimension to this order are 
\beq
\frac{\beta(g)}{2g}= -\frac{\beta_0}{2} \frac{\alpha_s}{4\pi} -\frac{\beta_1}{2} \left(\frac{\alpha_s}{4\pi}\right)^2, \qquad \gamma_m=6C_F\frac{\alpha_s}{4\pi} +  \left(3C_A^2 + \frac{97}{3}C_F C_A -\frac{10}{3}C_F n_f \right)\left(\frac{\alpha_s}{4\pi}\right)^2,
\eeq
where $\beta_0=\frac{11C_A}{3}-\frac{2n_f}{3}$ and $\beta_1=\frac{34}{3}C_A^2-2C_Fn_f -\frac{10}{3}C_An_f$. 
The two-loop evolution of the twist-two matrix elements reads \cite{Floratos:1981hs,Larin:1996wd}
\beq
\frac{\partial}{\partial \ln \mu} \begin{pmatrix} A^R_q \\ A^R_g \end{pmatrix}
= \left[ \frac{\alpha_s}{4\pi} X +\left(\frac{\alpha_s}{4\pi}\right)^2 Y\right] \begin{pmatrix} A^R_q \\ A^R_g\end{pmatrix} ,
\label{dglap}
\eeq
where 
\beq
X&=& \begin{pmatrix} -\frac{16C_F}{3} & \frac{4n_f}{3} \\ \frac{16C_F}{3} & -\frac{4n_f}{3} \end{pmatrix},  \\
Y&=&-2\begin{pmatrix} \frac{376}{27}C_FC_A  -\frac{112}{27} C_F^2 -\frac{104}{27} n_f C_F & 
-\frac{74}{27}C_Fn_f -\frac{35}{27}C_An_f \\
-\frac{376}{27}C_F C_A + \frac{112}{27}C_F^2 + \frac{104}{27}C_F n_f & \frac{74}{27}C_F n_f + \frac{35}{27}C_A n_f  \end{pmatrix}.
\eeq
This can be integrated as 
\beq
 \begin{pmatrix} A^R_q \\ A^R_g \end{pmatrix} = Z  \begin{pmatrix} A_q \\ A_g \end{pmatrix},
\eeq
where 
\beq
Z=1- \frac{X}{2}\frac{\alpha_s}{4\pi \epsilon} + \left(\frac{X^2}{8}+\frac{\beta_0X}{4}\right)\left(\frac{\alpha_s}{4\pi\epsilon} \right)^2 -\frac{Y}{2} \left(\frac{\alpha_s}{4\pi}\right)^2\frac{1}{2\epsilon}.
\eeq
For the renormalization constants, we now have  \cite{Tarrach:1981bi}
\beq
Z_F=1-\beta_0 \frac{\alpha_s}{4\pi \epsilon}  + \left(\beta_0\frac{\alpha_s}{4\pi\epsilon } \right)^2 - 2\beta_1\left(\frac{\alpha_s}{4\pi}\right)^2 \frac{1}{2\epsilon}
\eeq
and 
\begin{eqnarray}
Z_C& =&   
\frac{\alpha_s}{4\pi}
\frac{12 C_F}{\epsilon }
+\left(\frac{\alpha_s}{4\pi}\right)^2\left[
\frac{-12C_F\beta_0}{\epsilon
   ^2}+\frac{1}{\epsilon
   }\left(6 C_F^2+ \frac{194 C_A C_F}{3}-\frac{20 C_F n_f}{3}\right)
\right].
\end{eqnarray}
Using Mathematica, we have repeated the calculation in the previous section. The result for the renormalization constants is 
\begin{eqnarray}
Z_\psi   &=&  
1+ \frac{\alpha_s}{4\pi}\left( 
\frac{8 C_F}{3 \epsilon }
\right)
+\left(\frac{\alpha_s}{4\pi}\right)^2\left[
\frac{C_F \left(\frac{16 n_f}{9}-\frac{44 C_A}{9}\right)+\frac{32
   C_F^2}{9}}{\epsilon ^2}+\frac{C_F \left(\frac{188 C_A}{27}-\frac{52 n_f}{27}\right)-\frac{56
   C_F^2}{27}}{\epsilon }
\right]
,
\\
 Z_Q & =&   
 \frac{\alpha_s}{4\pi}\left( 
-\frac{2 n_f}{3 \epsilon }
\right)
+\left(\frac{\alpha_s}{4\pi}\right)^2\left[
\frac{\frac{11 C_A n_f}{9}-\frac{8 C_F n_f}{9}-\frac{4
   n_f^2}{9}}{\epsilon ^2}+\frac{-\frac{35}{54} C_A n_f-\frac{37 C_F n_f}{27}}{\epsilon }
\right]
,
\\
Z_T& =&   
1
+ \frac{\alpha_s}{4\pi}\left( 
\frac{2 n_f}{3 \epsilon }
\right)
+\left(\frac{\alpha_s}{4\pi}\right)^2\left[
\frac{-\frac{11 C_A n_f}{9}+\frac{8 C_F n_f}{9}+\frac{4
   n_f^2}{9}}{\epsilon ^2}+\frac{\frac{35 C_A n_f}{54}+\frac{37 C_F n_f}{27}}{\epsilon }
\right]
,
\\ 
Z_L& =&   
 \frac{\alpha_s}{4\pi}\left( 
-\frac{8 C_F}{3 \epsilon }
\right)
+\left(\frac{\alpha_s}{4\pi}\right)^2\left[
\frac{C_F \left(\frac{44 C_A}{9}-\frac{16
   n_f}{9}\right)-\frac{32 C_F^2}{9}}{\epsilon ^2}+\frac{C_F \left(\frac{52 n_f}{27}-\frac{188
   C_A}{27}\right)+\frac{56 C_F^2}{27}}{\epsilon }
\right]
,
\\ 
Z_M & =&   
 \frac{\alpha_s}{4\pi}\left( 
\frac{11 C_A}{12 \epsilon }
\right)
+\left(\frac{\alpha_s}{4\pi}\right)^2\left[
\frac{\frac{11 C_A n_f}{12}-\frac{121 C_A^2}{36}+\frac{2 C_F
   n_f}{9}}{\epsilon ^2}+\frac{-\frac{14 C_A n_f}{27}+\frac{17 C_A^2}{6}-\frac{5 C_F
   n_f}{108}}{\epsilon }
\right]
,
\\
Z_B& =&   
 \frac{\alpha_s}{4\pi}\left( 
-\frac{n_f}{6 \epsilon }
\right)
+\left(\frac{\alpha_s}{4\pi}\right)^2\left[
\frac{\frac{11 C_A n_f}{36}-\frac{2 C_F
   n_f}{9}-\frac{n_f^2}{9}}{\epsilon ^2}+\frac{-\frac{17}{54} C_A n_f-\frac{49 C_F
   n_f}{108}}{\epsilon }
\right]
,
\\
Z_S &=&
 \frac{\alpha_s}{4\pi}\left( 
-\frac{7 C_F}{3 \epsilon }
\right)
+\left(\frac{\alpha_s}{4\pi}\right)^2\left[
\frac{C_F \left(\frac{88 C_A}{9}-\frac{14 n_f}{9}\right)+\frac{8
   C_F^2}{9}}{\epsilon ^2}+\frac{C_F \left(\frac{11 n_f}{27}-\frac{406 C_A}{27}\right)-\frac{85
   C_F^2}{54}}{\epsilon }
\right]
,
\\
Z_K& =&   
 \frac{\alpha_s}{4\pi}\left( 
-\frac{2 C_F}{3 \epsilon }
\right)
+\left(\frac{\alpha_s}{4\pi}\right)^2\left[
\frac{C_F \left(\frac{11 C_A}{9}-\frac{4 n_f}{9}\right)-\frac{8
   C_F^2}{9}}{\epsilon ^2}+\frac{C_F \left(\frac{34 n_f}{27}-\frac{61 C_A}{54}\right)+\frac{2
   C_F^2}{27}}{\epsilon }
\right].
\end{eqnarray}

The anomaly coefficients in (\ref{ei}) and (\ref{ee}) are given by  
\begin{eqnarray}
x
& =&    \frac{\alpha_s}{4\pi}\left( 
\frac{n_f}{3}
\right)
+\left(\frac{\alpha_s}{4\pi}\right)^2\left[
\frac{17 C_A n_f}{27}+\frac{49 C_F n_f}{54}
\right],
\\
 y & =&   
 \frac{\alpha_s}{4\pi}\left( 
\frac{4 C_F}{3}
\right)
+\left(\frac{\alpha_s}{4\pi}\right)^2\left[
C_F \left(\frac{61 C_A}{27}-\frac{68 n_f}{27}\right)-\frac{4
   C_F^2}{27}
\right].
\end{eqnarray}
This leads to the main result of this paper, which is the two-loop version of (\ref{tgr})-(\ref{cons})
\begin{eqnarray}
 \eta_{\mu\nu}T^{\mu\nu}_{gR}&=&A_g^R(\mu)+4\bar{C}_g^R(\mu) \nonumber \\
&=& \frac{1}{2M^2}\left\langle P\right|\Biggl\{\frac{\alpha_s}{4\pi}\left( 
\frac{14}{3} C_F \left(m \bar{\psi }\psi  \right)_R-\frac{11}{6} C_A \left(F^2\right)_R
\right) 
 \label{tgtwoloop}  \\&&
+\left(\frac{\alpha_s}{4\pi}\right)^2\left[
 \left(C_F \left(\frac{812 C_A}{27}-\frac{22 n_f}{27}\right)+\frac{85
   C_F^2}{27}\right) \left(m \bar{\psi }\psi  \right)_R
   + \left(\frac{28 C_A n_f}{27}-\frac{17 C_A^2}{3}+\frac{5 C_F
   n_f}{54}\right)\left(F^2\right)_R
\right]
\Biggr\}\left| P\right\rangle , \nonumber
\end{eqnarray}
\begin{eqnarray}
\eta_{\mu\nu}T^{\mu\nu}_{qR} &=& A_q^R(\mu)+4\bar{C}_q^R(\mu) \nonumber \\
& =&   \frac{1}{2M^2}\left\langle P\right|\Biggl\{
\left(m \bar{\psi }\psi  \right)_R
+ \frac{\alpha_s}{4\pi}\left( 
\frac{4}{3} C_F \left(m \bar{\psi }\psi  \right)_R+\frac{1}{3} n_f
   \left(F^2\right)_R
\right) \label{tqtwoloop} \\ 
&& 
+\left(\frac{\alpha_s}{4\pi}\right)^2\left[
   \left(m \bar{\psi }\psi  \right)_R \left(C_F \left(\frac{61 C_A}{27}-\frac{68
   n_f}{27}\right)-\frac{4 C_F^2}{27}\right) 
   +\left(F^2\right)_R \left(\frac{17 C_A n_f}{27}+\frac{49 C_F
   n_f}{54}\right)
\right]
\Biggr\}\left| P\right\rangle , \nonumber 
\end{eqnarray}
and, similarly, (\ref{50}) now becomes
\begin{eqnarray}
&&A_q^R(\mu)+4\bar C_q^R(\mu)-\left( A_q+d \bar C_q \right)
\nonumber\\
& &= 
\frac{1}{2M^2}\left\langle P\right|\Biggl\{
 \frac{\alpha_s}{4\pi}\left( 
\frac{4  C_F}{3}\left(m\bar\psi \psi\right)_R+\frac{ n_f}{3}\left(F^2\right)_R
\right)
\nonumber\\&&
 \qquad +\left(\frac{\alpha_s}{4\pi}\right)^2\left[
   \left(\frac{17 C_A}{27}+\frac{49 C_F}{54}\right) n_f\left(F^2\right)_R+ \left(\frac{61 C_A
   C_F}{27}-\frac{68 C_F n_f}{27}-\frac{4 C_F^2}{27}\right)\left(m\bar\psi \psi\right)_R
\right]
\Biggr\}\left| P\right\rangle .
\end{eqnarray}
Moreover, Eq.~(\ref{rge3b}), together with the two-loop renormalization constants, 
leads to the two-loop evolution equation for the three-body operator,
\begin{eqnarray}
\frac{\partial}{\partial \ln \mu}
\left( g\bar\psi F^{\mu\nu}\gamma_\nu \psi \right)_R
& =&   
 \frac{\alpha_s}{4\pi}\left( 
 \left(-\frac{16 C_F}{3}-\frac{4 n_f}{3}\right)\left( g\bar\psi F^{\mu\nu}\gamma_\nu \psi \right)_R+\frac{4 
   C_F}{3}\partial^\mu\left( m \bar\psi \psi \right)_R
\right)
\nonumber\\&&
+\left(\frac{\alpha_s}{4\pi}\right)^2\left[
   \left(\frac{11 C_A}{18}+\frac{4 C_F}{9}\right) n_f \partial^\mu\left( F^2 \right)_R
   \right.
   \nonumber\\
   &&
   +\left(
   \left(\frac{20 C_F}{9}-\frac{70 C_A}{27}\right)n_f-\frac{752 C_A C_F}{27}+\frac{224
   C_F^2}{27}\right)\left( g\bar\psi F^{\mu\nu}\gamma_\nu \psi \right)_R 
  \nonumber\\
  &&\left.
    + \left(\frac{122 C_A C_F}{27}-\frac{136 C_F n_f}{27}-\frac{8
   C_F^2}{27}\right)\partial^\mu\left( m \bar\psi \psi \right)_R
\right] ,
\label{ee3}
\end{eqnarray}
extending the previous one-loop result (\ref{tot}).  
As mentioned below (\ref{full}), this two-loop result is needed to correctly evaluate the renormalization group evolution of $\bar{C}$. 
Let us  check the consistency bewteen  (\ref{ee3}) and (\ref{tgtwoloop}), (\ref{tqtwoloop}).
The scale dependence of $\bar C_q^R(\mu)$ at the two-loop accuracy may be calculated
by differentiating (\ref{tqtwoloop}) with respect to $\mu$, as
\beq
 \frac{\partial\bar{C}_q^R(\mu)}{\partial\ln \mu}&=&-\frac{1}{4}\left(
  \left[ \frac{\alpha_s}{4\pi} X +\left(\frac{\alpha_s}{4\pi}\right)^2 Y\right]_{qq} A^R_q (\mu)
  +\left[ \frac{\alpha_s}{4\pi} X +\left(\frac{\alpha_s}{4\pi}\right)^2 Y\right]_{qg}  A^R_g(\mu)
 \right)
%
\nonumber\\
&+&
\frac{1}{4}\frac{\partial\left(m \bar{\psi }\psi  \right)_R}{\partial\ln \mu}
+
 \frac{\alpha_s}{16\pi}\left( 
\frac{4}{3} C_F\frac{\partial \left(m \bar{\psi }\psi  \right)_R}{\partial\ln \mu}+\frac{1}{3} n_f
   \frac{\partial \left(F^2\right)_R}{\partial\ln \mu}
\right)
\nonumber\\
&+&\frac{\beta_R}{2g_R} \frac{\alpha_s}{4\pi}\left( 
\frac{4}{3} C_F \left(m \bar{\psi }\psi  \right)_R+\frac{1}{3} n_f
   \left(F^2\right)_R
\right)
,
\label{tqtwoloopd}
\end{eqnarray}
where we have substituted (\ref{dglap}) for $\partial A^R_q(\mu)/\partial\ln \mu$, and
$\partial \alpha_s/\partial\ln \mu=4(\beta_R/2g_R)\alpha_s$ from the definition of the $\beta$ function.
The remaining $\mu$-derivative terms are determined by the renormalization group equations which directly follow from (\ref{o2ren}) and (\ref{o4ren})
\beq
\frac{\partial \left(F^2\right)_R}{\partial\ln \mu}&=&
\frac{\alpha_s}{4\pi}\left[ 
 \left(\frac{22 C_A}{3}-\frac{4 n_f}{3}\right)\left(F^2\right)_R-24  C_F\left(m \bar{\psi }\psi  \right)_R
\right]
\nonumber\\&&
+\left(\frac{\alpha_s}{4\pi}\right)^2\left[
\left(-\frac{40 C_A
   n_f}{3}+\frac{136 C_A^2}{3}-8 C_F n_f\right)\left(F^2\right)_R 
\right.\nonumber\\&&\left.   
  \qquad \qquad+ \left(C_F \left(\frac{80 n_f}{3}-\frac{776 C_A}{3}\right)-24
   C_F^2\right)\left(m \bar{\psi }\psi  \right)_R
\right]
, \label{rgeff}
\\
\frac{\partial\left(m \bar{\psi }\psi  \right)_R}{\partial\ln \mu}&=&0, 
\label{rgempsi}
\eeq 
 whose solution is given by (\ref{mass}). 
We then eliminate $A_{q,g}^R(\mu)$ from (\ref{tqtwoloopd}) using  (\ref{tqtwoloop}) and (\ref{tgtwoloop}).
The resulting equation exactly coincides with the one obtained by taking the nonforward matrix element of  (\ref{ee3}).

On the other hand, in the forward limit $\Delta=0$ we have an additional constraint, $A_q^R(\mu)+A_g^R(\mu)=1$. Using this and  (\ref{tqtwoloop}), we can eliminate $A_{q,g}^R$ from
 (\ref{tqtwoloopd}) 
and find\footnote{ If we instead use $A_q^R(\mu) \to 1- A_g^R(\mu)$ and  (\ref{tgtwoloop}) in  (\ref{tqtwoloopd}), 
 we obtain  (\ref{tqtwoloopdc}) up to extra terms that vanish with the use of the relation (\ref{mass}).}
\beq
 \frac{\partial\bar{C}_q^R(\mu)}{\partial\ln \mu}&=&
 \frac{\alpha_s}{4\pi}\left[ 
\bar{C}_q^R(\mu) \left(-\frac{16 C_F}{3}-\frac{4 n_f}{3}\right)-\frac{n_f}{3}+ \left(\frac{4
   C_F}{3}+\frac{n_f}{3}\right)\frac{\left\langle P \right| \left(m \bar{\psi }\psi  \right)_R \left| P \right\rangle}{2M^2}
\right]
\nonumber\\&&
+\left(\frac{\alpha_s}{4\pi}\right)^2\Biggl[
\bar{C}_q^R(\mu) \left(C_F \left(\frac{20 n_f}{9}-\frac{752
   C_A}{27}\right)-\frac{70 C_A n_f}{27}+\frac{224 C_F^2}{27}\right)
   \nonumber\\
   &&\left.   
  -\frac{35}{54} C_A n_f-\frac{37 C_F n_f}{27} +
   \left(\frac{4 C_F n_f}{9}+\frac{n_f^2}{9}\right)\frac{\left\langle P \right| \left(F^2  \right)_R \left| P \right\rangle}{2M^2}
      \right.
   \nonumber\\
   &&
   +\left(C_F \left(\frac{122
   C_A}{27}-\frac{5 n_f}{3}\right)+\frac{35 C_A n_f}{54}-\frac{8 C_F^2}{27}\right)\frac{\left\langle P \right| \left(m \bar{\psi }\psi  \right)_R \left| P \right\rangle}{2M^2} 
\Biggr].
\label{tqtwoloopdc}
\end{eqnarray}
Finally, we use (\ref{mass}) 
to eliminate $\left\langle P \right| \left(F^2  \right)_R \left| P \right\rangle$, yielding
\beq
 \frac{\partial\bar{C}_q^R(\mu)}{\partial\ln \mu}&=&
 \frac{\alpha_s}{4\pi}\Biggl[
\bar{C}_q^R(\mu) \left(-\frac{16 C_F}{3}-\frac{4 n_f}{3}\right)-\frac{n_f}{3}-\frac{n_f}{\beta_0}\left( \frac{8 C_F}{9}+\frac{2
   n_f}{9}\right) 
   \nonumber\\
   &&
   + \left(\frac{n_f}{\beta_0}\left( \frac{8 C_F}{9}+\frac{2
   n_f}{9}\right) +\frac{4 C_F}{3}+\frac{n_f}{3}\right)\frac{\left\langle P \right| \left(m \bar{\psi }\psi  \right)_R \left| P \right\rangle}{2M^2}
\Biggr]+\cdots .
\label{tqtwoloopdcb}
\end{eqnarray}
In the chiral limit, the solution of this equation approaches (\ref{asymptc}) asymptotically.

The above derivation makes it clear that the $\mu$-dependence of $\bar{C}_{q,g}^R$ is completely fixed by the condition $T^{\mu\nu}=T^{\mu\nu}_R$ and the anomalous dimension of the twist-two matrix elements $A_{q,g}^R$. In view of this, the RG equation  $\partial \bar{C}^R_{q,g}/\partial \ln \mu=\cdots$ is somewhat redundant and can be even misleading as the naive counting in $\alpha_s$ does not work. We actually know the explicit solution of this equation  including the integration constants, see  (\ref{con})-(\ref{cons}) and (\ref{tgtwoloop})-(\ref{tqtwoloop}).

Even more explicit formulas can be derived by  using the well-known expression 
\beq
A_q^R\left(\mu \right)=\frac{n_f}{4
   C_F+n_f}+\frac{4 C_F A_q^R\left(\mu _0\right)+n_f \left(A_q^R\left(\mu _0\right)-1\right)}{4 C_F+n_f}\left(\frac{\alpha _s\left(\mu
   \right)}{\alpha _s(\mu_0 )}\right)^{\frac{ 8 C_F+2n_f}{3 \beta_0}}+\cdots\ ,
\eeq
with a certain starting scale $\mu_0$. Here, the ellipses denote the next-to-leading contributions associated 
with $\beta_1$ and $Y$ of (\ref{dglap}), namely, the order $\alpha_s^2$ contributions when expanded in the power
series in $\alpha_s$.
Substituting (\ref{mass}) into (\ref{tqtwoloop}) to eliminate $\left\langle P \right| \left(F^2  \right)_R \left| P \right\rangle$,
we obtain
\beq
\bar{C}_q^R(\mu)&=&
- \frac{1}{4} \left( \frac{n_f}{4C_F+n_f} + \frac{2n_f}{3\beta_0}\right) + \frac{1}{4}\left(\frac{2 n_f}{3 \beta _0}+1\right)\frac{\left\langle P \right| \left(m \bar{\psi }\psi  \right)_R \left| P \right\rangle}{2M^2}
   \nonumber\\
   &&
   -\frac{4 C_F A_q^R\left(\mu _0\right)+n_f \left(A_q^R\left(\mu _0\right)-1\right)}{4(4 C_F+n_f)}\left(\frac{\alpha _s\left(\mu
   \right)}{\alpha _s(\mu_0 )}\right)^{\frac{ 8 C_F+2n_f}{3 \beta_0}}
    \nonumber\\
   &&
+ \frac{\alpha_s(\mu)}{4\pi}\left[
\frac{n_f \left(-\frac{34
   C_A}{27}-\frac{49 C_F}{27}\right)}{4\beta _0}+\frac{ \beta_1
   n_f}{6 \beta _0^2}
   \right.
    \nonumber\\
   &+&\left.
    \frac{1}{4}\left(\frac{n_f \left(\frac{34 C_A}{27}+\frac{157
   C_F}{27}\right)}{\beta _0}+\frac{4 C_F}{3}-\frac{2 \beta_1 n_f}{3 \beta _0^2}\right)\frac{\left\langle P \right| \left(m \bar{\psi }\psi  \right)_R \left| P \right\rangle}{2M^2}
\right]+\cdots
,
\eeq
which reproduces (\ref{asymptc}), now taking into account the quark mass effect.
Numerically, we have
\beq
\left. \bar{C}_q^R(\mu)\right|_{n_f=3}&\simeq&-0.146
-0.25 \left(A_q^R\left(\mu _0\right)-0.36\right)\left(\frac{\alpha _s\left(\mu \right)}{\alpha _s(\mu_0 )}\right)^{\frac{50}{81}} -0.01 \alpha _s(\mu
   )
\nonumber\\&&
+
\left( 0.306+ 0.08 \alpha _s(\mu )\right)
\frac{\left\langle P \right| \left(m \bar{\psi }\psi  \right)_R \left| P \right\rangle}{2M^2},
\label{nu1}
\eeq
and
\beq
\left. \bar{C}_q^R(\mu)\right|_{n_f=2}&\simeq&-0.103
-0.25 \left(A_q^R\left(\mu _0\right)-0.27\right)\left(\frac{\alpha _s\left(\mu \right)}{\alpha _s(\mu_0 )}\right)^{\frac{44}{87}}  -0.004
   \alpha _s(\mu )
\nonumber\\&&
+
\left(0.284+ 0.061 \alpha _s(\mu )\right)\frac{\left\langle P \right| \left(m \bar{\psi }\psi  \right)_R \left| P \right\rangle}{2M^2}.
\label{nu2}
\eeq
Thus, the important correction comes from the evolution of the twist-two form factor $A_q^R$,
while the other corrections  play a minor ($\sim$a few percent) role.

\section{Conclusions}

In this paper we have studied the renormalization of the QCD trace anomaly separately for the quark and gluon parts of the energy momentum tensor.  While the renormalization of the total anomaly $T=T_q+T_g$ is well understood in the literature \cite{Tarrach:1981bi}, our analysis at the quark and gluon level has revealed some interesting new features. The bare and renormalized $(T_{q,g})^\alpha_\alpha$ differ by finite operators, and this difference can be systematically computed order by order in $\alpha_s$. It is interesting to notice that, at one loop,  the renormalized $T_q$ gives the $n_f$ part of the beta function. However,  this  property no longer holds at two-loop, see (\ref{tgtwoloop}). Besides, the partition of the total anomaly can be different if one uses other regularization schemes (see, e.g., the `gradient flow' regularization \cite{Makino:2014taa}), and it is interesting to study their mutual relations.  We have also found that $\bar{C}_{q,g}(\mu)$ does not go to zero as $\mu \to \infty$ even in the chiral limit, contrary to what one would naively expect from the one-loop calculation (\ref{full}). 

Our result has interesting phenomenological implications. In  \cite{Hatta:2018ina}, the relation between $F^2$ and $(T_{q,g})^\alpha_\alpha$ has been worked out for the bare quantities. If a more careful analysis reveals that one should use the renormalized relation, the numerical result in \cite{Hatta:2018ina} may have to be revised.
Another place where $\bar{C}_{q,g}$ plays a role is the nucleon's transverse spin sum rule. It has been shown in \cite{Hatta:2012jm,Leader:2012ar,Chakrabarti:2015lba} (see also \cite{Ji:2012vj}) that Ji's sum rule \cite{Ji:1996ek} does not hold for a transversely polarized nucleon unless the nucleon is at rest. One has, for the quark/gluon total angular momentum $J_{q,g}$,  
\beq
J_{q,g}=\frac{1}{2}(A_{q,g}+B_{q,g}) + f(P_z)\bar{C}_{q,g}
\eeq
where $f(P_z)$ is a frame-dependent function (depends on the nucleon longitudinal momentum $P_z$) which vanishes at $P_z=0$ and approaches $\frac{1}{2}$ as $P_z\to \infty$. Asymptotically,  $\frac{1}{2}(A_q+B_q) = 0.18$ while $\bar{C}_q=-\bar{C}_g \approx -0.15$ for $n_f=3$, so the effect of the last term can be actually quite significant.

\section*{Acknowledgments}
We thank Maxim Polyakov for insightful remarks which triggered this work and Volodya Braun for discussion.  
This material is based on work supported by the U.S. Department of Energy, Office of Science, Office of Nuclear Physics, under Contracts No. DE- SC0012704.

\end{document}